# DIRECT MEASUREMENT CAPABILITIES OF IN SITU WATER DENSITY

A.N. Grekov, N.A. Grekov

Institute of Natural and Technical Systems, Sevastopol

*E-mail: oceanmhi@ya.ru*

*Taking into account the new approach to the thermodynamics of seawater, it is proposed to carry out direct measurements of density and temperature in situ. The possibilities of vibration, hydrostatic and acoustic density meters are considered. The design of a new acoustic density meter is proposed and an algorithm for its operation is presented.*

The developed and created classical automated CTD systems for measuring the most important physicochemical characteristics of seawater, apparently, have exhausted their capabilities.

The fact is that oceanographic science puts forward new requirements for the knowledge of sea water. This is the study of deep waters over 2000 m and shelf zones of oceans and seas, where the salt composition is significantly different from the classically accepted composition of seawater, which was determined for 135 samples collected in different regions of the oceans in the layer from the surface to 100 m [1].

The created automated CTD systems measure temperature, electrical conductivity and pressure. By electrical conductivity, taking into account the correction for temperature and pressure, using the equation of state of seawater (MUS-80), practical salinity is determined, i.e. practical salinity depends only on conductivity [2].

However, the thermodynamic properties of seawater are directly affected by the mass of various substances dissolved in it, and this mass per unit volume is defined as absolute salinity.

The advantages of determining the absolute salinity are quite fully reflected in the IOC UNESCO document dated April 27, 2009 in paragraph 4.3.2.1.

The IOC recommended using a new approach to the thermodynamics of seawater and in practice to start implementing the "Thermodynamic equation of state of seawater-2010" - abbreviated as TEOS-10. TEOS-10 uses absolute salinity instead of practical salinity [3].

From the point of view of automation of measurements, the question arises of how to measure the absolute salinity using modern advances in instrumentation.

In practice, many methods have been developed to measure the absolute salinity, the origins of which date back to 1818. Without listing all the measurement methods, we will make a reference to these methods, which are published in [4]. These methods are not suitable for direct in situ salinity measurements. The most promising methods in oceanography to meet the TEOS-10 requirement are direct measurements of density and temperature in situ, and then, if necessary, the salinity is calculated from these two parameters.

Therefore, in this work, we will consider methods for measuring density that are suitable for in situ.

**Vibration density meter** [5] is often used for continuous density measurements and consists of a tubular resonator, inside which the investigated water flows. The tube material should have a low internal damping and a high modulus of elasticity. For precision measurements, quartz glass can be used.

The square of the angular resonance frequency of the transverse vibrations of the tube is described by the equation

$$\omega^2 = \frac{rEJ}{ml^3}, \tag{1}$$

where $r$ – is a constant that depends on the conditions for fixing the tube; $E$ – modulus of elasticity of the tube material; $J$ – moment of inertia of the cross-section of the tube; $m$ и $l$ – tube weight and length.

Taking into account the mass of the liquid inside the tube, the equation can be transformed to the form

$$\omega^2 = \frac{k}{m + \rho V}, \tag{2}$$

where $\rho$ и $V$ – density and volume of the liquid; $k$ – system constant, determined when calibrating the meter.

From equation (2) it is easy to obtain the value of the density of the liquid

$$\rho = \left(\frac{k}{\omega^2} - m\right) / V. \tag{3}$$

Experimental studies have shown that the oscillation frequency of tubular resonators does not depend on the pressure and viscosity of the flowing liquid, but depends only on the mass of the liquid inside the resonator. However, a change in the temperature of the environment and the monitored liquid causes the appearance of an additional error associated with a change in the geometric dimensions of the sensor, its modulus of elasticity and the density of the wall material. All these parameters of the sensor must be monitored and corrected accordingly when measuring density.

O. Kratky, H. Leopold and H. Stabinger [6] used this method to create a density meter with a digital indicator in 1969. This instrument allowed to achieve the accuracy $\pm 1{,}5 \cdot 10^{-6}$.

Klaus Kremling [7] studied the applicability of this instrument in oceanography. For this, 9 samples were taken in the salinity range from 9 to 39 ‰ at six temperatures from 0 to 25 °C and compared with the equation obtained by R.A. Cox, M.J. McCartney and F. Culkin [8] using the hydrostatic balance method. The density values obtained by the two methods are in good agreement with each other within the absolute error of $0{,}008\,\sigma_t$.

The known method of hydrostatic balance [9], which has several varieties and consists in the fact that liquids with different densities $\rho_0$ and $\rho_1$, which are in inverted vessels, in which the pressure is below atmospheric pressure, will be at different heights $h_0$ and $h_1$. Having an exemplary density of $\rho_0$, the density of the unknown liquid is determined from the difference in heights of $h_1$ and $h_2$. However, this method, like the buoyancy method based on the Archimedes principle, is used mainly in laboratory conditions.

Radioisotope methods [10] to measure the density of a liquid use the attenuation of the penetrating radiation of gamma rays through the analyzed liquid. The error of this method reaches 1–2%. Besides, the hardware implementation is very cumbersome. Therefore, it is problematic to use radioisotope methods to measure density in situ at the present stage.

Consider a promising method for measuring water density using an acoustic signal. This method can be used on the open sea and under high vibrations and pressures.

It is known that the specific acoustic resistance of an infinite medium is defined as the product of the density and the speed of sound propagation in the medium

$$Z = \pm\, \rho c\,. \qquad (4)$$

Let the plane wave fall normally on the plane boundary $x = 0$ (Figure 1) between two homogeneous media with densities $\rho_1$ and $\rho_2$, speeds of sound $c_1$ and $c_2$, wave impedance $z_1$ and $z_2$, pressure amplitudes: $A_\text{И}$ – radiation, $A_\text{T}$ – transmission, $A_\text{R}$ – reflection.

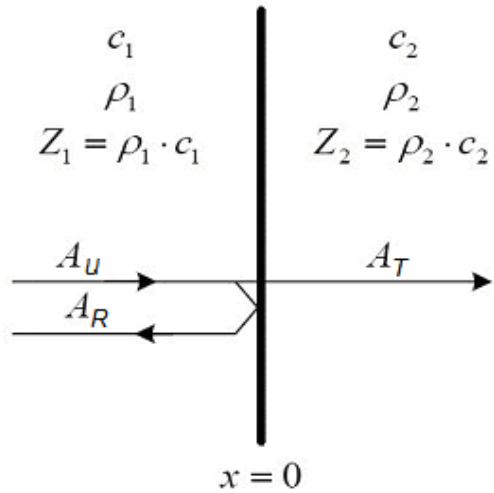

Figure 1 – Scheme of reflection and transmission acoustic signal at the border of two media

Consider the Fresnel formula for the reflection coefficient $R$

$$R = \frac{\rho_2 c_2 - \rho_1 c_1}{\rho_2 c_2 + \rho_1 c_1} = \frac{z_2 - z_1}{z_2 + z_1} \qquad (5)$$

and transmission coefficient $T$

$$T = \frac{2 \rho_2 c_2}{\rho_2 c_2 + \rho_1 c_1} = \frac{2 z_2}{z_2 + z_1}. \qquad (6)$$

These equations can be used to determine the density of water if the other variables are known.

For example, for a simple density measurement method, using a buffer material with a known velocity $c_1$ and a density $\rho_1$, which separates the acoustic transducer and the medium under study with a known velocity $c_2$, it is possible to determine the density of water $\rho_2$:

$$\rho_2 = \frac{\rho_1 c_1 (1+R)}{c_2 (1+R)}. \qquad (7)$$

Using relative wave impedance

$$W = \frac{\rho_2 c_2}{\rho_1 c_1}. \qquad (8)$$

The coefficients of reflection and transmission through the relative characteristic impedance are defined as

$$R = \frac{W-1}{W+1}, \quad T = \frac{2W}{W+1}. \tag{9}$$

The amplitude of the radiated sound pressure $A_u$ (Fig. 1) decomposes into the pressure amplitude of reflection $A_R$ and transmission $A_T$, then $R$ can be defined as

$$R = \frac{A_R}{A_u} \tag{10}$$

and the quantity $T$ is defined as

$$T = \frac{A_T}{A_H}. \tag{11}$$

Based on this principle, an ultrasonic meter for the speed of sound and acoustic impedance [11] was developed and created, which is used to calculate the density. This device was used for measurements on Lake Merseburg-Ost, located in Central Germany [12].

Interesting results were obtained from direct in situ density measurements that differed from those obtained with CTD probes. Therefore, in situ density measurements by ultrasonic methods are promising for studying the density structure in natural aqueous media.

The authors of the article have developed a design of an acoustic water density transducer and an algorithm for its operation (Figure 2). The acoustic transducer is manufactured from a stable material with a low coefficient of thermal expansion.

Structurally, it consists of a monolithic rectangle, on one of the sides of which a piezoelectric transducer is glued. The transducer is sealed against the ingress of sea water. The plane of the rectangle opposite to the transducer is in contact with air with a characteristic impedance of $Z_в$, which allows the condition $Z_в \ll Z_1$ to be fulfilled.

The transducer has a rectangular groove through which water flows, the resistivity and the speed of sound to be determined. An electrical impulse is supplied to the piezoelectric transducer from the microcontroller, where it is converted into an acoustic signal. The same piezoelectric transducer is a receiver of acoustic signals reflected from the walls, which, after conversion, go to the inputs of the ADC and TDC of the microcontroller.

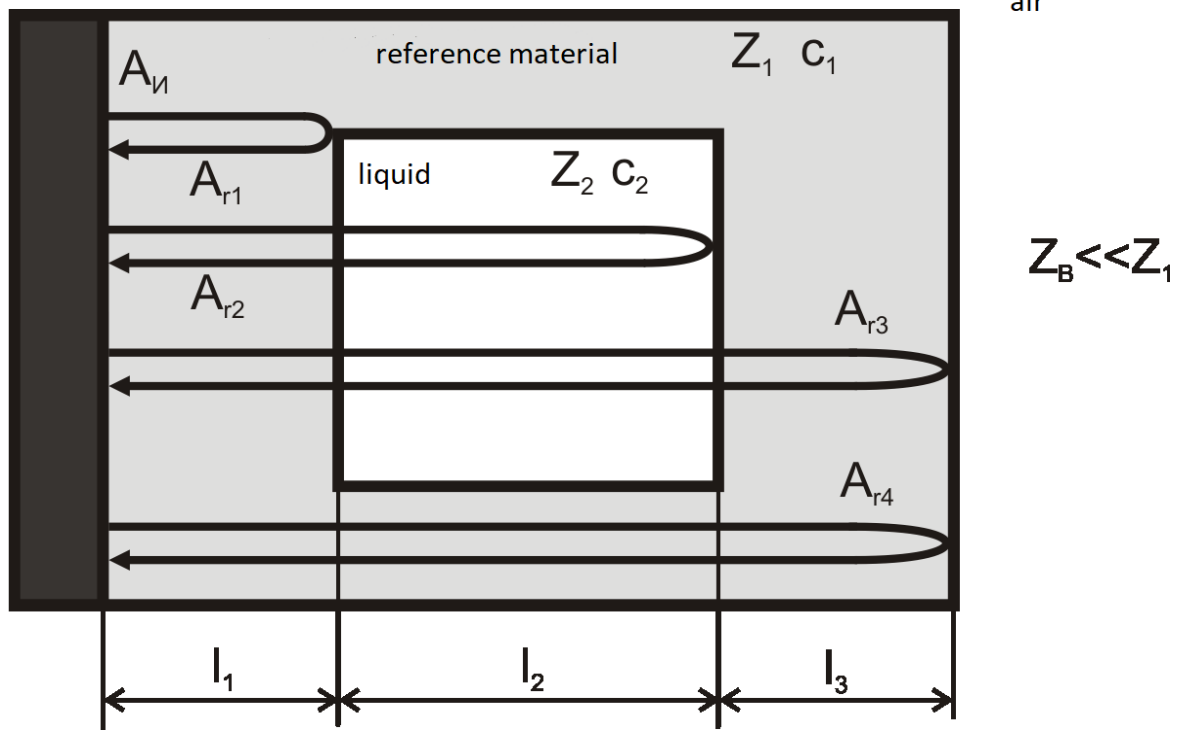

Figure 2 – Diagram of the transducer device and the path of acoustic pulses

$l_1$ – - the distance between the piezoelectric transducer and the groove wall;

$l_2$ – the distance between the opposite walls of the groove;

$l_3$ – the distance between the wall of the groove and the border of the wall, which is in contact with air;

$Z_1$ и $c_1$ – wave impedance and speed of sound of a stable material;

$A_И$ – radiation amplitude;

$A_{r1}$, $A_{r2}$, $A_{r3}$, $A_{r4}$, – amplitudes of reflection from various walls;

$Z_2$ и $c_2$ – wave impedance and speed of sound of measured water;

$\rho$ – density of the determined water.

**Algorithm of the converter.** First, using an exemplary liquid with known $Z_{образц}$, $\rho$ and $c$, determine the density of a stable material $\rho_1$.

$$\rho_1 = \frac{Z_{образц}}{c_1} \cdot \frac{1-R}{1+R} . \tag{12}$$

The value of $R$ is determined from the reflection amplitudes $A_{r1}$ and $A_{r4}$

$$R = -\frac{A_{r1}}{A_{r4}} \cdot \frac{1}{K}, \tag{13}$$

where $K$ is the constant of the converter.

The sound speed of a stable material $c_1$ is defined as

$$c_1 = \frac{2(l_2 + l_3)}{t_{r4} - t_{r1}}. \tag{14}$$

The calculated and measured values are used to determine

$$Z_2 = \rho_1 c_1 \frac{1+R}{1-R}. \tag{15}$$

The magnitude of the speed of sound in water $c_2$ can be determined

$$c_2 = \frac{l_2}{t_{r2} - t_{r1}}. \tag{16}$$

Knowing the values of $Z_2$ and $c_2$, we determine the density of water $\rho_2$

$$\rho_2 = \frac{Z_2}{c_2}. \tag{17}$$

This article does not specify a specific distance $l_1$, $l_2$, $l_3$, they are calculated from the properties of the stable material used.

The proposed density conversion device, in contrast to the known ones, allows you to obtain additional values of $A$, $R$, $C$ both for liquid and for reference materials, in fact, this sensor implements a multichannel parametric measuring system, which in turn makes it possible to increase the measurement accuracy due to compensation instability of sensor elements as a result of destabilizing factors.